# Reinforcement learning enabled the design of compact and efficient integrated photonic devices

**Mirbek Turduev[1], Emre Bor[2], Onur Alparslan[3], Y. Sinan Hanay[4], Hamza Kurt[5], Shin'ichi Arakawa[3], and Masayuki Murata[3]**

[1]Department of Electrical and Electronics Engineering, Kyrgyz Turkish Manas University, Bishkek 720042, Kyrgyzstan
[2]Department of Electrical and Electronics Engineering, TOBB University of Economics and Technology, Ankara 06560, Turkey
[3]Graduate School of Information Science and Technology, Osaka University, Osaka 565-0871, Japan
[4]Department of Computer Engineering, Akdeniz University, Antalya 07070, Turkey
[5]School of Electrical Engineering, Korea Advanced Institute of Science and Technology (KAIST), Daejeon, 34141, Republic of Korea

E-mail: mirbek.turduev@manas.edu.kg

**Abstract**

In this paper, we introduce the design approach of integrated photonic devices by employing reinforcement learning known as attractor selection. Here, we combined three-dimensional finite-difference time-domain method with attractor selection algorithm, which is based on artificial neural networks, to achieve ultra-compact and highly efficient photonic devices with low crosstalk such as wavelength demultiplexers and a polarization splitter. The presented devices consist of silicon-on-insulator materials, which are compatible with complementary metal-oxide-semiconductor technology, and their structural dimensions enable the possible fabrication process in the future. The numerical results are presented for the near-infrared wavelengths at around 1550 nm, and the performance of designed photonic devices with footprint of 3×3 µm$^2$ are compared with the previously reported structures. Consequently, the reinforcement learning is successfully applied to design smaller and superior integrated photonic devices where the use of presented approach can be further expanded to different applications.

Keywords: integrated photonic devices, wavelength demultiplexer, polarization beamsplitter, machine learning, artificial neural networks, reinforcement learning

## 1. Introduction

Design of photonic devices are, in general, based on physical knowledge and intuition where small number of structural parameters are finely tuned by hand. Nevertheless, brute-force designs are incompetent to reveal efficient photonic devices. In the last decade, novel design approaches introduced which mainly integrate optimization algorithms with numerical methods to form remarkable photonic devices by genuinely calculating the interaction between light and matter. In this manner, various photonic devices are presented such as subwavelength focusing lenses, mode order converters, asymmetric light transmitters and optical cloaks [1–6]. Evidently, these approaches took the design of photonic devices one step further.

In recent years, the so-called "inverse-design" methods, which use mostly silicon-on-insulator (SOI) materials, are introduced to design ultra-compact, highly efficient and complementary metal-oxide-semiconductor (CMOS)-compatible integrated photonic devices. For instance, an approach based on convex optimization is presented to design several wavelength demultiplexers [7–10]. Also, direct-binary search algorithm is applied to design a polarization beamsplitter and a polarization rotator [11, 12].

Lately, machine learning and artificial neural networks have attracted great interest from researchers and enabled design of photonic devices in a different manner. For example, deep learning accelerated the design of an all-dielectric metasurface structures [13, 14]. Also, Bragg grating devices are obtained by training artificial neural networks [15]. In addition, a machine learning algorithm is proposed for focusing and optical coupling of light [16]. Moreover, a reinforcement learning is applied to design optical coupler and asymmetric light transmitter devices [17].

In the present study, we demonstrate the design of wavelength demultiplexers and a polarization beamsplitter operating at near-infrared wavelengths via attractor selection (AttSel) algorithm which is considered as reinforcement learning and based on artificial neural networks [18, 19]. For this purpose, we combined the algorithm and three-dimensional (3D) finite-difference time-domain (FDTD) method [20]. The designed structures exhibit high optical performance in an ultra-compact area and their numerical investigations are presented in detail. It should be noted that fabrication constraints are considered throughout the design process [11] which enables the possible fabrication of devices in the future. We believe that the proposed approach is not only restricted to design integrated photonic devices, but it may lead the advances in different photonic designs.

## 2. Design approach and numerical investigation of integrated photonic devices

Even though photonic integrated circuits have superior features comparing to integrated electronic circuits, they have a main deficiency of lower integration density [21, 22]. On the other hand, there is a trade-off between optical performance and structural dimensions of a photonic device. For this reason, designing a highly efficient photonic structure with small footprint is a challenge which can be overcome by applying advanced search algorithms or machine learning.

In general, machine learning is divided into two branches such as supervised learning and unsupervised learning. In these branches, simply, a data set is considered to train and test the learning algorithm where data are independently collected, i.e., the algorithm does not have any effect on during the collection process of data. On the other hand, artificial neural networks (ANNs) are considered as the subclass of machine learning methods which is powerful for modelling nonlinear relationships.

In a very recent study, ANNs are applied to characterize the cases of "forward" and "inverse" designs [15]. An ANNs is applied to map four structural parameters of Bragg gratings to two optical characteristics. In forward modelling, the selected four structural parameters are considered as input to ANNs whereas the two optical characteristics are used as output. In inverse modelling, the input and output parameters are switched. In other words, two optical values are inserted as inputs to ANNs whereas four structural values are introduced as outputs. As a result, as expected, ANNs easily found the nonlinear relationship between small number of input and output parameters. In similar cases, ANNs is applied for a regression problem which is considered as a branch of supervised learning.

However, in the studies of inverse design of photonic devices, hundreds of structural parameters are optimized to find only one or small number of optical characteristics [1–12]. In this case, for inverse modelling, a simple regression method based on ANNs would not be able to find the characteristics between large number of structural parameters and small number of output values. The reason of this possible failure is that the small number of input parameters may not be informative enough for ANNs to map them to large number of output parameters. Moreover, in the case of inverse modelling, if a data set does not contain any results of good optical characteristics, ANNs would not predict structural parameters for a desired optical performance since ANNs do not operate as search/optimization algorithms.

To overcome this issue, another branch of machine learning known as reinforcement learning would be reasonable to apply for photonic designs. In the reinforcement learning, the algorithm works during the sampling of data set which differs from supervised and unsupervised learning. Therefore, a reinforcement learning algorithm can find the values of large number of structural parameters for even small number of desired optical characteristics. For this reason, in our recent work, we have applied AttSel algorithm which is a reinforcement learning method, to design optical coupler and asymmetric light transmitter [17].

AttSel models the interaction of the metabolic reaction network and the gene regulatory network in a cell [18]. The cell growth requires converting the nutrition in the environment by metabolic reactions of proteins to the substances necessary for the growth. The proteins that carry out this conversion are produced by the gene reaction network, where each gene has an expression level for controlling the protein production level. The expression levels change with the rate of substance production. If the substance production rate is high, it implies that the conditions are favorable, so the state is saved as an attractor and the expression levels do not change much from the attractor. Otherwise, the cell tries to adapt to the environment by searching for a new set of expression levels that can increase the substance production rate. The expression levels start to deviate from the last attractor by the impulse of noise, which becomes dominant. This deviation allows the cell to try different expression levels until it finds a new set of expression levels that make the conditions favorable again.

The individual cells in the photonic device can be interpreted as the proteins, whose expression levels are represented as $x = (x_1, x_2, \cdots, x_n)$. If the expression level of $x_i$, which is between [-1, 1], is $x_i < 0$, then the $i^{th}$ individual cell in the photonic device is defined to be an air. Otherwise, it is defined to be Si.

The analytical expression for calculation of $x_i$ is [19]:

$$\frac{dx_i}{dt} = \alpha \cdot \left( f\left(\sum_j W_{ij} \cdot x_j - \theta\right) - x_i \right) + \eta. \quad (1)$$

In Eq. (1), $\alpha$ is the growth rate, which increases when the performance of the tested photonic device increases. $\eta$ is the Gaussian noise. If the performance of the tested photonic device is high, the system moves in the proximity of an attractor. Otherwise, the noise becomes dominant and the expression levels change randomly until a new attractor state is found. $W$ is the regulatory matrix, which stores the attractors. It is a Hopfield neural network [23]. An orthogonal projection is used for calculating $W$ [24, 25]. Multiplying $W$ with $x$ moves the $x$ to an attractor. $\theta$ is the threshold value of expression level. The deterministic behavior is regulated by a sigmoidal function

$$f(z) = tanh(\mu z), \quad (2)$$

where, $\mu$ is the gain parameter. The performance of the tested photonic device is expressed in terms of error $E$. Then $E$ is converted to the growth rate $\alpha$ by

$$\alpha = \frac{1}{1+exp(\delta \cdot (E-\zeta))}, \quad (3)$$

where $\delta$ is the gradient multiplier and the $\zeta$ is the threshold of error rate. The attractor selection algorithm parameters were empirically set to $\delta = 0.20$ and $\mu = 10$. $\zeta$ is set to the error of the last found attractor. In general, the variance of the noise $\eta$ is set to 0.3. In case the system is stuck at an attractor without improvement in the transmission rate for a long time, $\eta$ is increased to 1 temporarily so that the system can search for other attractors. The attractor selection algorithm works in an iterative manner. In each iteration, there are three phases. Firstly, the transmission levels of the photonic device are calculated by FDTD method. In the second phase, AttSel calculates the new expression levels $x = (x_1, x_2, \cdots, x_n)$ by using these transmission levels. According to the new expression levels, the structure of the photonic device is updated in the third phase.

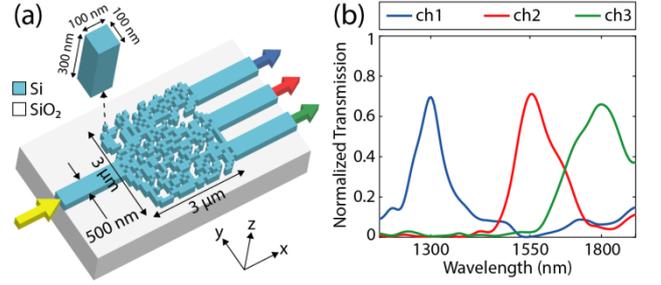

**Figure 1.** (a) Perspective view of designed device for demultiplexing the wavelengths of 1300 nm, 1550 nm, 1800 nm) with its structural parameters and (b) plots of normalized transmission efficiencies at the output waveguides. Same colors are used for arrays at output waveguides and corresponding calculated transmission efficiencies.

The computational experiments are done by running nine computers in parallel fashion to speed-up the numerical calculations. The CPUs in the computers are 2× Intel Xeon E5-2697 v2, 2× Intel Xeon E5-2643 v4, Intel i7-6700K, Intel i7-6700K, Intel i7-6700K, 2× Intel Xeon E5-2690 v4, 2× Intel Xeon X7560, 2× Intel Xeon E5-2690, and Intel i7-3960x.

In this study, by applying AttSel, we designed photonic structures with sizes of 3×3 µm² and slab thickness of 300 nm which are composed of Si material on a SiO$_2$ substrate. The design area is divided into square pixels with dimensions of 100×100 nm² where the algorithm determined the states of each pixel to be Si or air according to the desired optical property [17]. Here, the refractive indices of Si and air are fixed as $n_{Si}$=3.46 and $n_{air}$=1.0, respectively, at near-infrared wavelengths around 1550 nm. Input and output waveguides having width of 500 nm are connected to design area to form a photonic device. It should be noted that these waveguides can only support fundamental modes of both transverse-electric (TE) and transverse-magnetic (TM) polarizations of light.

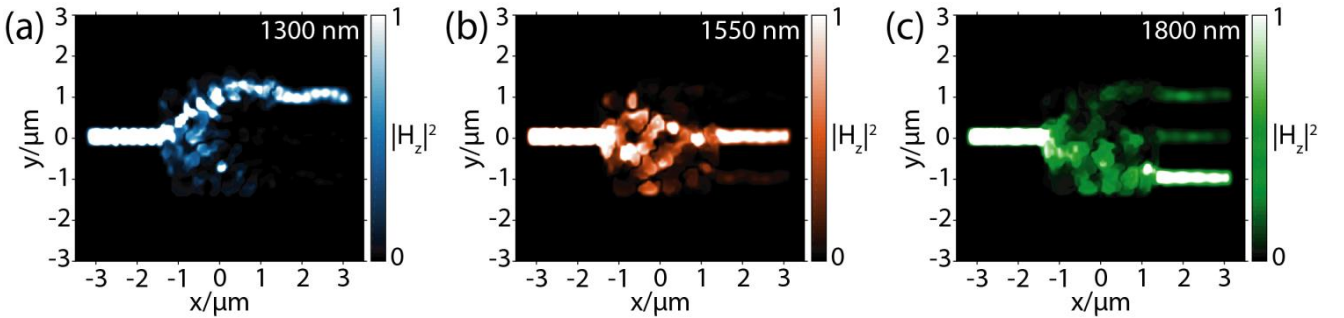

**Figure 2.** Calculated magnetic field intesity distributions at the wavelengths of (a) 1300 nm, (b) 1550 nm and (c) 1800 nm for the firstly designed wavelengths demultiplexer device.

Firstly, we present the wavelength demultiplexer which separates the incident wavelengths of 1300 nm, 1550 nm and 1800 nm to different output waveguides. Here, guided-mode $TE_0$ is considered as incident light in 3D FDTD simulations. The algorithm tried to increase the transmission efficiency of a desired wavelength in a selected waveguide. Also, we aimed to reduce the transmission of undesired wavelengths in a waveguide to prevent crosstalk between the output waveguides. The corresponding photonic device is schematically represented in Figure 1(a). The distance between adjacent output waveguides is fixed to 500 nm.

The transmission efficiencies are calculated as 70% (–1.55 dB), 71% (–1.48 dB) and 66% (–1.80 dB) at the selected wavelengths of 1300 nm, 1550 nm and 1800 nm, respectively, which are plotted in Figure 1(b). As can be seen, there exist negligible level of crosstalk which is under –12 dB and points out efficient wavelength demultiplexing. It can be concluded that main loss of input optical power is due to out-of-plane radiation.

Additionally, magnetic field intensity ($|H_z|^2$) distributions are calculated at selected wavelengths and given in Figure 2. Here, the incident light at wavelength of 1300 nm is directed to the upper output waveguide whereas wavelength of 1800 nm is canalized to lower output waveguide. The middle output waveguide conducts the wavelengths at around 1550 nm.

In wavelength division multiplexing (WDM) systems, compactness, channel spacing, number of channels, high transmission efficiency and low crosstalk are highly desirable criteria. For this reason, we applied AttSel to fulfil all the requirements in a wavelength demultiplexer device. Here, we designed another photonic device to separate closely selected wavelengths of 1500 nm, 1550 nm and 1600 nm with transmission efficiencies of 75% (–1.25 dB) at different output waveguides. The designed photonic devices are schematically represented in Figure 3(a) with its structural parameters. Also, the corresponding transmission efficiencies are plotted in Figure 3(b). As can be seen from this figure, a low crosstalk of under –12 dB occurs between selected wavelengths at output waveguides. It should be noted that the corresponding device is excited by fundamental $TE_0$ mode in numerical calculations. In addition, we extracted the magnetic field distributions at the selected wavelengths which are presented in Figure 4. It is clear to see that the proposed device is capable of demultiplexing the selected wavelengths with small channel spacing of 50 nm.

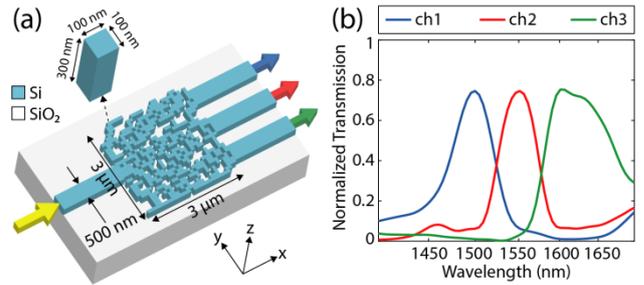

**Figure 3.** (a) Perspective view of designed device for demultiplexing the wavelengths of 1500 nm, 1550 nm, 1600 nm) with its structural parameters and (b) plots of normalized transmission efficiencies at the output waveguides. Same colors are used for arrays at output waveguides and corresponding calculated transmission efficiencies.

In a recent study, the proposed wavelength demultiplexer can separate three near-infrared wavelengths between 1500 nm and 1580 nm with channel spacing of 40 nm. The corresponding device has numerically calculated transmission efficiencies varying between –1.68 dB and –1.35 dB around and covers an area of 24.75 $\mu m^2$[8]. On the other hand, our secondly designed wavelength demultiplexer device given in Figure 3 is capable of separating three wavelengths between 1500 nm and 1600 nm with channel spacing of 50 nm. Also, our device has ultra-compact sizes of 9.0 $\mu m^2$ and transmission efficiencies –1.25 dB. When comparing these two wavelength demultiplexers, our structure has fairly smaller footprint and higher transmission efficiencies for a slightly larger channel spacing of 10 nm. In addition, our structure has low crosstalk between waveguides even if the waveguides are closely placed to each other with a distance of 500 nm.

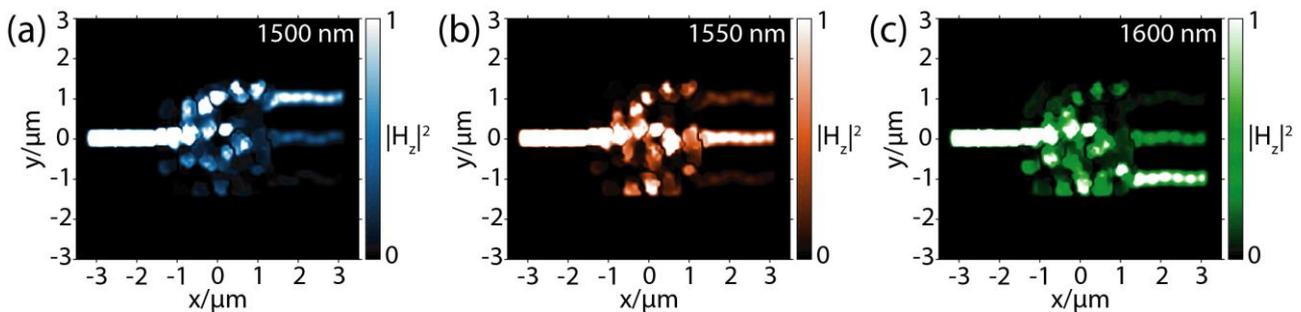

**Figure 4.** Calculated magnetic field intesity distributions at the wavelengths of (a) 1500 nm, (b) 1550 nm and (c) 1600 nm for the secondly designed wavelengths demultiplexer device.

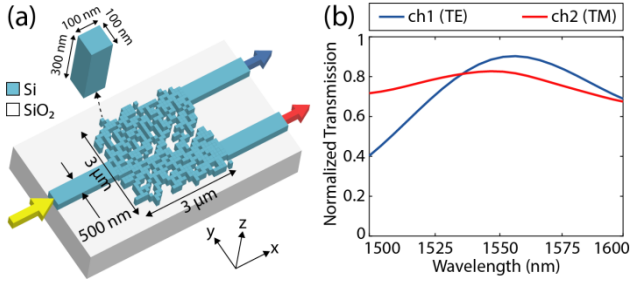

**Figure 5.** (a) Perspective view of designed device for seperating TE and TM polarizations of light at wavelength of 1550 nm with its structural parameters. (b) Plots of normalized transmission efficiencies at the output waveguides. Same colors are used for arrays at output waveguides and corresponding calculated transmission efficiencies.

In photonic integrated circuits, SOI systems are usually preferred due to high refractive index contrast between Si and $SiO_2$ materials. Because of this difference, birefringence emerges, and operation of designed structures become polarization-dependent. Thus, integrated photonic devices are designed by restricting their operation to one polarization of light. Therefore, compact and efficient polarization beamsplitters are crucial requirements for photonic integrated circuits. For this reason, we designed a polarization beamsplitter via AttSel. The proposed photonic device routes TM and TE polarizations of light to different output waveguides for the wavelength of 1550 nm. In Figure 5(a), the proposed device is represented. In this device, the incident light of TE polarization is canalized to the upper waveguide whereas the TM polarization is coupled to the bottom waveguide. It should be noted that the output waveguides are separated by a distance of 1500 nm.

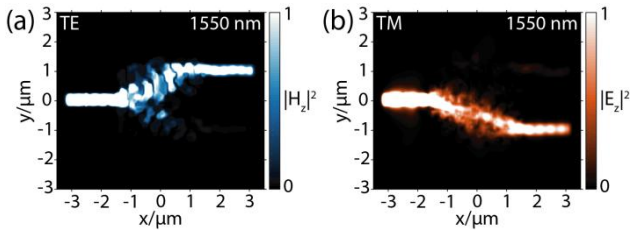

**Figure 6.** Calculated (a) magnetic field intesity distribution of TE polarization and (b) electric field intesity distribution of TM polarization at the wavelength of 1550 for the presented polarization beamsplitter.

The calculated transmissions efficiencies are equal to 89% (–0.5 dB) and 83% (–0.8 dB) for TE and TM polarizations, respectively, which are plotted in Figure 5(b). Also, the extinction ratios are calculated as 15.9 dB and 21.2 dB, respectively, for TE and TM polarizations. The bandwidth for transmission efficiency over 70% and extinction ratio greater than 13.44 dB is emerged as 73 nm (4.65%). In addition, crosstalk is calculated as –22.0 dB for the output waveguide which guides TE polarization whereas it is equal to –16.3 dB for the output waveguide of TM polarization.

In terms of optical performance, the designed photonic device with footprint of 9.0 $\mu m^2$ outperforms the previously reported polarization splitter with sizes of 5.76 $\mu m^2$ [11]. Furthermore, we calculated the magnetic field intensity distribution ($|H_z|^2$) of TM polarization at wavelength of 1550 nm which is given in Figure 6(a). The calculated electric field intensity ($|E_z|^2$) distributions of TE polarization are extracted at the design wavelength of 1550 nm and represented in Figure 6(b).

## 3. Conclusion

In conclusion, we present the application of a reinforcement learning algorithm, which is based on attractor selection mechanism, to design SOI-based integrated photonic devices such as wavelength demultiplexers and a polarization beamsplitter for near-infrared wavelengths. Therefore, 3D FDTD method is integrated into AttSel algorithm for photonic designs. The proposed devices have ultra-compact sizes and performs fairly well for the desired optical property. The evaluation of numerical results is presented in detail. Throughout the design process, fabrication constraints are considered which enables the possible fabrication and characterization of the presented photonic devices in the future. We believe that the introduced reinforcement learning algorithm has the potential to pave the way for designing remarkable photonic devices.

## Acknowledgements

This work was supported by the Scientific and Technological Research Council of Turkey (TUBITAK) under Project 116F182.